\documentclass[twocolumn,trackchanges]{aastex701}

\usepackage{soul}

\begin{document}

\newcommand{\oii}{\mathrm{[O\,\textsc{ii}]}\,\lambda3726}
\newcommand{\oiii}{\mathrm{[O\,\textsc{iii}]}}
\newcommand{\OIIIaur}{\oiii\,\lambda4363} 
\newcommand{\oiiiA}{\oiii\,\lambda4959}
\newcommand{\oiiiB}{\oiii\,\lambda5007}
\newcommand{\nii}{\mathrm{[N\,\textsc{ii}]}\,\lambda6584}
\newcommand{\sii}{\mathrm{[S\,\textsc{ii}]}\,\lambda\lambda6717,6731}
\newcommand{\hb}{\mathrm{H}\beta}
\newcommand{\ha}{\mathrm{H}\alpha}

\newcommand{\nmosfiresamp}{64}

\title{Metal-Poor Gas Accretion Drives Giant Clump Formation at $0.6 < z < 2.6$}

\author[orcid=0000-0003-0780-9526,gname=Visal, sname=Sok]{Visal Sok}
\affiliation{Department of Astrophysical and Planetary Sciences, University of Colorado, 2000 Colorado Ave, Boulder, CO 80309, USA}
\affiliation{Department of Physics and Astronomy, York University, 4700 Keele Street, Toronto, ON, M3J 1P3, Canada}
\email[show]{visal.sok@colorado.edu}  

\author[orcid=0000-0002-9330-9108,gname=Adam, sname='Muzzin']{Adam Muzzin} 
\affiliation{Department of Physics and Astronomy, York University, 4700 Keele Street, Toronto, ON, M3J 1P3, Canada}
\email[no]{muzzinad@yorku.ca}  

\author[orcid=0000-0001-6003-0541]{Ben Forrest} 
\affiliation{Department of Physics and Astronomy, University of California, Davis, One Shields Avenue, Davis, CA 95616, USA}
\email[no]{test}

\author[orcid=0000-0002-6572-7089]{Gillian Wilson} 
\affiliation{Department of Physics, University of California, Merced, 5200 North Lake Road, Merced, CA 92543, USA}
\email[no]{test}  

\author[orcid=0009-0006-2702-1962]{Jialu Chen}
\affiliation{Department of Astronomy, Harvard University, 60 Garden Street, Cambridge, MA 02138, USA}
\email[no]{test}

\author[orcid=0000-0002-3503-8899]{Vivian Yun Yan Tan}
\affiliation{Department of Physics and Astronomy, York University, 4700 Keele Street, Toronto, ON, M3J 1P3, Canada}
\email[no]{tanvivia@yorku.ca}

\author[orcid=0009-0000-8716-7695]{Sunna Withers} 
\affiliation{Department of Physics and Astronomy, York University, 4700 Keele Street, Toronto, ON, M3J 1P3, Canada}
\email[no]{sunnaw@yorku.ca}

\author[]{Roberto Abraham}
\affiliation{David A. Dunlap Department of Astronomy and Astrophysics, University of Toronto, 50 St. George Street, Toronto, Ontario, M5S 3H4, Canada}
\affiliation{Dunlap Institute for Astronomy and Astrophysics, 50 St. George Street, Toronto, Ontario, M5S 3H4, Canada}
\email{roberto.abraham@utoronto.ca}

\author[0000-0001-5984-0395]{Maru\v{s}a Brada\v{c}}
\affiliation{Faculty of Mathematics and Physics, Jadranska ulica 19, SI-1000 Ljubljana, Slovenia}
\affiliation{Department of Physics and Astronomy, University of California Davis, 1 Shields Avenue, Davis, CA 95616, USA}
\email{marusa.bradac@fmf.uni-lj.si}

\author[0000-0001-8489-2349]{Vicente Estrada-Carpenter}
\affiliation{School of Earth and Space Exploration, Arizona State University, Tempe, AZ 85287, USA}
\affiliation{Beus Center for Cosmic Foundations, Arizona State University, Tempe, AZ 85287, USA}
\email{vestrad9@asu.edu}

\author[0000-0001-9298-3523]{Kartheik G. Iyer}
\affiliation{Columbia Astrophysics Laboratory, Columbia University, 550 West 120th Street, New York, NY 10027, USA}
\affiliation{Center for Computational Astrophysics, Flatiron Institute, 162 Fifth Avenue, New York, NY 10010, USA}
\email{kiyer@flatironinstitute.org}

\author[0000-0003-3243-9969]{Nicholas S. Martis}
\affiliation{Faculty of Mathematics and Physics, Jadranska ulica 19, SI-1000 Ljubljana, Slovenia}
\email{nicholas.martis@fmf.uni-lj.si}

\author{Ga\"el Noirot}
\affiliation{Space Telescope Science Institute, 3700 San Martin Drive, Baltimore, Maryland 21218, USA}
\email{gnoirot@stsci.edu}

\author[0000-0001-8830-2166]{Ghassan T. E. Sarrouh}
\affiliation{Department of Physics and Astronomy, York University, 4700 Keele Street, Toronto, ON, M3J 1P3, Canada}
\email{gsarrouh@yorku.ca}

\author[0000-0002-7712-7857]{Marcin Sawicki}
\affiliation{Department of Astronomy and Physics and Institute for Computational Astrophysics, Saint Mary's University, 923 Robie Street, Halifax, Nova Scotia B3H 3C3, Canada}
\email{marcin.sawicki@smu.ca}

\author[0000-0002-4201-7367]{Chris J. Willott}
\affiliation{National Research Council of Canada, Herzberg Astronomy \& Astrophysics Research Centre, 5071 West Saanich Road, Victoria, BC, V9E 2E7, Canada}
\email{chris.willott@nrc.ca}

\author[orcid=0000-0002-9466-2763]{Aur\'elien Henry} 
\affiliation{Department of Physics, University of California, Merced, 5200 North Lake Road, Merced, CA 92543, USA}
\email[no]{}

\author[orcid=0009-0009-9848-3074]{Naadiyah Jagga} 
\affiliation{Department of Physics and Astronomy, York University, 4700 Keele Street, Toronto, ON, M3J 1P3, Canada}
\email[no]{}

\author[orcid=0000-0001-9002-3502]{Danilo Marchesini} 
\affiliation{Department of Physics and Astronomy, Tufts University, 574 Boston Avenue, Suite 304, Medford, MA 02155, USA}
\email[no]{danilo.marchesini@tufts.edu}

\author[orcid=0000-0002-2446-8770]{Ian McConachie} 
\affiliation{Department of Astronomy, University of Wisconsin-Madison, 475 N. Charter St., Madison, WI 53706 USA}
\email[no]{}

\author[0009-0009-2307-2350]{Katherine Myers}
\affiliation{Department of Physics and Astronomy, York University, 4700 Keele Street, Toronto, ON, M3J 1P3, Canada}
\email{kjmyers@yorku.ca}

\author{Nelson Nunes} 
\affiliation{Department of Physics and Astronomy, York University, 4700 Keele Street, Toronto, ON, M3J 1P3, Canada}
\email[no]{}

\author[0000-0002-6265-2675]{Luke Robbins} 
\affiliation{Department of Physics and Astronomy, Tufts University, 574 Boston Avenue, Suite 304, Medford, MA 02155, USA}
\email[no]{}

\begin{abstract}

The physical properties of kiloparsec-scale clumps in high-redshift star-forming galaxies (SFGs) contain crucial constraints on how they assemble. Building on recent work that indicates the presence of a metallicity offset in clumpy galaxies compared to nonclumpy SFGs, we analyze the chemical abundance in a large sample of ${\sim}300$ SFGs between $0.6<z<2.6$ using LEGA-C, MOSDEF and CANUCS near-infrared spectroscopic observations. We find that clumpy galaxies generally have lower gas-phase metallcities compared to the mass-metallicity relation, while nonclumpy galaxies have higher metallicities. We further investigate the relationship between the resolved stellar properties of clumps and the integrated gas-phase metallicity of the host galaxies using a subset of galaxies observed in the CANUCS fields. In particular, clumps in SFGs with metallicity below the mass-metallicity relation (i.e., $\Delta Z \leq 0$) are generally younger and have higher SFRs, compared to clumps whose host galaxies have $\Delta Z > 0$. We do not find a significant mass difference between these two clump populations. Finally, we compute the merger statistic using the Gini-M20 morphological parameters and find that the majority of clumpy galaxies are not classified as mergers based on their stellar mass maps. The results suggest that the clumpy nature of cosmic noon galaxies is linked to metal-poor gas accretion events that trigger star formation and dilute metallicities. 

\end{abstract}



\section{Introduction} 

It is well known that high-redshift star-forming galaxies generally host kiloparsec-scale clumpy structures \citep[e.g.,][]{Wuyts2012, Guo2012, Kalita2025a, Kalita2025b, Zhu2026, Chugunov2026}. In fact, clumpy galaxies becomes increasing ubiquitous at higher-$z$, with the fraction of clumpy galaxies following a similar redshift evolution to that of the cosmic star formation density \citep[e.g.,][]{Shibuya2016, Sok2022, Sattari2023, Vega2025, Sok2025b}. 
Furthermore, observations from the James Webb Space Telescope of distant galaxies show that clumpy substructures persist at $z>5$, enabled by its sensitivity and by gravitational lensing. These observations allow us to resolve both large kiloparsec-scale star-forming clumps and smaller parsec-scale stellar structures \citep[e.g.,][]{Messa2024, Adamo2024, Mowla2024, Fujimoto2025, Claeyssens2023, Claeyssens2025, Bradac2025}.
The unique mode of clumpy star formation at higher redshifts suggests that galaxy assembly history is inherently clumpy, with even Milky Way-like galaxies potentially hosting clump formation in the past (e.g., \citealt{Mowla2024, Tan2024}). This further raises questions on whether clumpy star formation represents a dominant mode of galaxy growth at early epochs, and how clumps are resilient to feedback such as supernovae explosions, galactic winds, and ISM turbulence. Understanding the physical mechanisms that drive clump formation and how these processes evolve over cosmic time are therefore paramount for understanding galaxy evolution. 

Despite the abundance of star-forming clumps at cosmic noon, their origin and evolution are relatively unconstrained. The common framework for clump formation is violent disk instability (VDI), in which clumps form due to gravitational instabilities in gas-rich regions \citep[e.g.,][]{Toomre1964, Dekel2009a}. The gas-rich nature of these disks is attributed to the continuous accretion from cosmological gas streams \citep[e.g.,][]{Keres2005, Keres2009, Dekel2009a, Dekel2009b, Voort2012, Nelson2013}. The accreting gas is expected to be metal-poor \citep{Voort2012}, in which clump metallicities are expected to be lower than their host disk, as observed in simulations \citep{Ceverino2016}. 
A result in the increase of the gas mass fraction due to accretion is the formation of massive clumps that could resist destructive mechanisms such as shearing and feedback (e.g., \citealt{Fensch2021, Renaud2024}). 
Observational constraints for clump formation via VDI generally probe the Toomre $Q$ parameter of the gas disks \citep[e.g.,][]{Forster2011, Genzel2011, Girard2018, Fujimoto2025}. However, such argument relies on simplified assumptions about disk structure. Future studies therefore need to consider more complex formalisms of the Toomre stability theory to better predict disk stability \citep[e.g.,][]{Romeo2010, Romeo2011, Bacchini2024}. 

Another pathway for clump formation is through mergers, in which gas is compressed by tidal forces arising from galaxy interactions \citep[e.g.,][]{Renaud2015, Nakazato2024}. Direct observational support comes from kinematic studies of smaller samples of galaxies, including \cite{Puech2010, Menendez2013, Calabro2019}. Indirect inferences about clump formation resulting from mergers are also made by using the redshift evolution of the merger rate to explain the redshift evolution of the fraction of clumpy galaxies \citep{Guo2015, Vega2025}. Clumps may further be of \textit{ex-situ} origins, that is, as being accreted satellites as observed in simulations \citep[e.g.,][]{Mandelker2014, Mandelker2017} and suggested in observations \citep[e.g.,][]{Ribeiro2017, Zanella2019}. 

Addressing the mechanisms driving clump formation requires observational data that link clump formation to a physical process, such as gas inflow. 
Recently, in \cite{Sok2025}, we showed that in a sample of approximately 30 star-forming galaxies, those that are clumpy have a metallicity offset of $\sim -0.1$ dex compared to nonclumpy galaxies. Similarly, resolved metallicity maps for approximately $20$ SFGs obtained with JWST NIRISS slitless grism data show that at least half of clump populations have a metallicity offset of $-0.1$ dex relative to the host galaxy disk \citep{Estrada2025}. While the sample size of these studies is small ($<30$), these findings suggest that clump formation can occur as a result of metal-poor gas accretion events.

Building upon these previous studies, this paper aims to investigate the relationship between clump formation and gas-phase metallicity. We first present new gas-phase metallicity measurements for ${\sim}70$ SFGs in the Canadian NIRISS Unbiased Cluster Survey (CANUCS, \citealt{Willott2022}) fields, using MOSFIRE observations. Combined with the unparalleled multi-wavelength imaging in the CANUCS data set, we examine the relationship between the global gas-phase metallicity of host galaxies and the resolved properties of star-forming clumps, derived from spectral energy distribution (SED) fitting. In addition, we compile existing spectroscopic data from the Large Early Galaxy Astrophysics Census (LEGA-C) survey \citep{vandelWel2016} and MOSFIRE Deep Evolution Field (MOSDEF) survey \citep{Kriek2015} to further place constraints the relationship between gas-phase metallicities and clumpiness. This multi-wavelength approach enables us to directly test the physical processes driving clump formation at cosmic noon. 

The paper is organized as follows. Section \ref{sec:ch3_data} describes the data used in this work. We describe the methodology in Section \ref{sec:ch3_method}. More specifically, the section explains how clumpy and nonclumpy galaxies are identified, and how the stellar properties of the clumps are estimated. We discuss the results in Section \ref{sec:ch3_results} and Section \ref{sec:ch3_metallicity_analyses} explores the gas-phase metallicity of SFGs and its relation to galaxy clumpiness. Section \ref{sec:ch3_clump_properties} expands on this by examining the relationship between metallicity offset with respect to the mass-metallicity relation and properties of star-forming clumps. We discuss the implications for clump formation mechanisms in Section \ref{sec:ch3_discussion} and summarize our work in Section \ref{sec:ch3_conclusion}.  

Throughout this paper, the term metallicity represents the gas-phase oxygen abundance, and is quoted as $12+\log(\mathrm{O/H})$.
All magnitudes are expressed in the AB system \citep{Oke1983}. We adopt a $\Lambda$CDM cosmological model of the universe with $\Omega_\lambda$ = 0.7, $\Omega_\mathrm{M}$ = 0.3, and a Hubble constant of $H_0$ = 70 km/s/Mpc. 

\section{Data} \label{sec:ch3_data}

To provide an in-depth analysis of the relationship between galaxy clumpiness and gas-phase metallicity across a range of redshifts, we make use of several extragalactic surveys.

We present new near-infrared spectroscopic observations of galaxies in the Canadian NIRISS Unbiased Cluster Survey (CANUCS, \citealt{Willott2022}) fields, obtained with the MOSFIRE instrument \citep{McLean2012} on the 10-meter Keck I telescope. At the time of the MOSFIRE observations, CANUCS was among the first extragalactic fields with deep, multi-wavelength JWST/NIRCam imaging. The CANUCS data further enable robust measurements of the properties of clumps including stellar masses and ages. 
In addition, we combine high-quality emission line measurements from the Large Early Galaxy Astrophysics Census (LEGA-C; \citealt{vandelWel2016}) survey and MOSFIRE Deep Evolution Field (MOSDEF; \citealt{Kriek2015}) survey. These datasets are chosen for their large galaxy samples and availability of ancillary imaging, enabling us to investigate the connection between galaxy morphology and their gas-phase metallicity over a broad redshift range. 

In the following sections, we describe the spectroscopic and photometric data used in this work, starting with our targeted observations in the CANUCS fields, followed by the LEGA-C and MOSDEF surveys.

\subsection{CANUCS feat. MOSFIRE Observations}
The primary focus of this study is to understand the relationship between the properties of star-forming clumps and the global gas-phase metallicity of their host galaxies. 

We make use of the JWST CANUCS fields due to its rich NIRCam and NIRISS broad-band and medium-band imaging for five cluster fields; namely Abell 370, MACS-J0416, MACS-J0417, MACS-J1149, and MACS-J1423. The data also includes the addition of NIRCAM parallel imaging of a flanking field for each cluster. These images reach 3$\sigma$ AB magnitude depths of 27–30.
The detailed list of available filters in each field is provided in \cite{Sarrouh2025}. The addition of medium-band imaging specifically enables more accurate modelling of the strong nebular emission observed in high-redshift galaxies (e.g., \citealt{Withers2023}). This leads to better photometric redshift determinations and stellar mass estimates, compared to solely relying on broad-band JWST data, which often leads to underestimated stellar masses (e.g., \citealt{Sarrouh2024}). 

Building on our previous work, where we investigated the clump mass and age distributions using JWST CANUCS imaging \citep{Sok2025b}, we now extend the analysis by inferring gas-phase metallicities from the $\nii/\ha$ ratio using Keck MOSFIRE observations. While JWST/NIRSpec prism observations exist for these fields, the spectral resolution is insufficient to resolve the $\nii/\ha$ ratio at these redshifts This allows us to explore the overall relationship between clump properties and the global properties of their host galaxies, such as gas-phase metallicity.

\subsubsection{Observations and Data Reduction}

We obtained a total of four nights on the Keck I telescope with MOSFIRE to perform deep spectroscopic observations of galaxies in the CANUCS fields. These observations spanned from February 2024 to January 2025. We targeted galaxies whose ${\ha}$ and ${\nii}$ emission lines fall within the wavelength coverage of the H ($1.450-1.826~\mu m$) and K ($1.450-1.826~\mu m$) bands. 
This corresponds to a redshift interval of 1.3-1.7 and 2.0-2.6 for ${\ha}$.
In total, we created 17 (7 H and 10 K) masks using the MAGMA slit mask design software. We adopt a slit width of $0^{\prime\prime}.7$, corresponding to a spectral resolution of $R\sim3600$.

The primary targets were selected from the catalogue presented in \cite{Sarrouh2025}. We determined slit-placement priorities for each mask based on several criteria. In particular, we targeted only star-forming galaxies and prioritized those that are massive and have higher SED-based $\ha$ flux estimates.  
The priority on stellar mass ensures that both $\ha$ and $\nii$ could be observed with high S/N, particularly at higher redshifts where metallicity is expected to be lower. The $\ha$ estimates were obtained from the SED-derived star formation rate, with a dust correction applied based on the galaxy's SED-inferred $A_V$. Since most primary targets lack prior spectroscopic observations, we relied on photometric redshifts to ensure that both $\ha$ and $\nii$ fall within the H/K bands. When available, we crossmatched our targets with the CANUCS NIRISS WFSS catalogue (Noirot et al., in prep.) to determine whether $\ha$ was available, and rejected sources whose $\ha$ line falls on skylines. 
During the first two half nights, we included galaxies with stellar masses greater than $10^{9} M_\odot$, and SED-based $\ha$ fluxes greater than $1\times 10^{-18}~\mathrm{erg~s^{-1}~cm^{-2}}$, maximizing the number of targets on slit. However, our observing strategy was changed for the rest of the observations to only target galaxies with $F_\mathrm{H\alpha} \geq 5\times 10^{-18}~\mathrm{erg~s^{-1}~cm^{-2}}$ as a result of non-detection of galaxies with lower $\ha$ fluxes within the integration time of around 1.5 to 2.5 hours.


We followed a similar observing strategy as the MOSDEF survey, specifically using an ABA$^\prime$B$^\prime$ dither pattern with nods of $+1^{\prime\prime}.5$, $-1^{\prime\prime}.2$, $+1^{\prime\prime}.2$, $-1^{\prime\prime}.5$, and individual exposures of 2 and 3 minutes for the H and K bands, respectively. In total, we placed 201 slits on primary targets, and successfully measured spectroscopic redshifts for 127 of them based on the visual identification of $\ha$ unaffected by skylines. The overall detection rate of $\ha$ for galaxies on slit is limited by the first few masks observed that included targets with lower SED-derived ${\ha}$ fluxes.

\begin{figure}[!t]
    \centering
    \includegraphics[width=\linewidth]{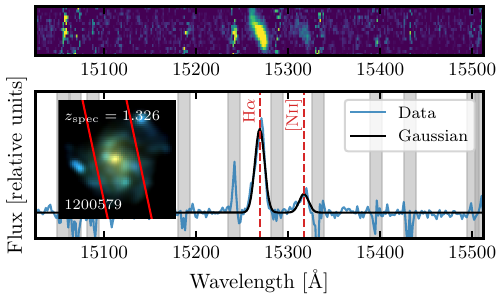}
    \includegraphics[width=\linewidth]{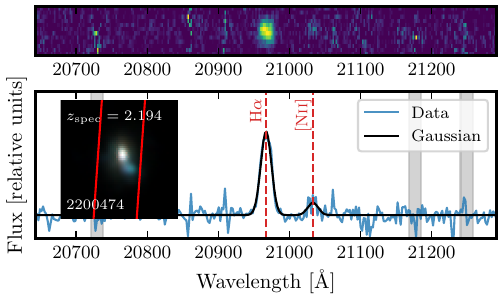}
    \includegraphics[width=\linewidth]{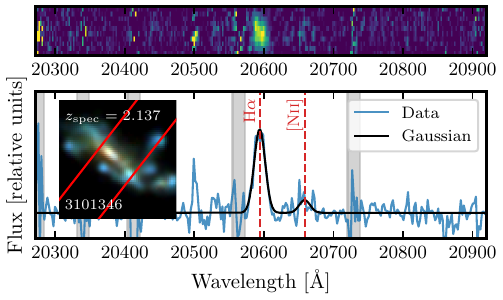}
    \caption{Examples of 2D spectra and extracted 1D spectra of CANCUS galaxies. Inset figures show the composite NIRCAM image of the galaxy, with the red line denoting the MOSFIRE slit position angle and width. The dark grey regions denote wavelengths affected by strong skylines. }
    \label{fig:spectra_example}
\end{figure}

We used the custom IDL pipeline of the MOSDEF survey, described in \cite{Kriek2015}, to reduce the data and produce two-dimensional (2D) science spectra for each slit on the mask. The one-dimensional (1D) science and error spectra were extracted based on an optimal extraction algorithm \citep{Horne1986}. Figure \ref{fig:spectra_example} shows examples of the reduced spectra. 
We do not apply any correction for slit losses, as this would be a multiplicative factor applied to both $\ha$ and $\nii$ fluxes, assuming weak metallicity gradients. However, we note that the metallicity gradient may differ from one galaxy to another. 

\subsection{The LEGA-C Survey}

We utilize the third data release of LEGA-C \citep{vandelWel2021}. LEGA-C is a European Southern Observatory (ESO) Public Spectroscopic survey of approximately 3500 galaxies at $0.6 < z < 1$, observed with the Very Large Telescope/Visible Multi-Object Spectrograph (VIMOS; \citealt{Fevre2003}) at a spectral resolution of $R\sim3500$. Galaxies are located in the Cosmic Evolution Survey (COSMOS; \citealt{Scoville2007}) field, and were selected as star-forming galaxies from the UltraVISTA \citep{McCracken2012} K-band catalogue presented by \cite{Muzzin2013}. 

We further cross-match the LEGA-C emission line flux catalogue with deconvolved ground-based imaging of the COSMOS field \citep{Sok2022}. The deconvolved imaging is available for over 20,000 star-forming galaxies between $0.5<z<2$, with wavelength coverage spanning from the Subaru $\mathrm{B_j}$-band to the UltraVISTA Ks-band. In this paper, we use these images to construct the inferred rest-frame $U$ surface brightness maps from SED fitting for the LEGA-C galaxies and classify them as being clumpy or nonclumpy (see \S \ref{sec:ch3_method}).

\subsection{The MOSDEF Survey}

We also include data from the MOSDEF survey, a 47-night observational program with MOSFIRE that targeted $\sim1500$ galaxies at $1.37<z<3.80$. Specifically, the survey provided medium-resolution rest-frame optical spectra for these galaxies, including bright emission lines such as ${\ha}$ and ${\nii}$. The galaxies are located in three of the CANDELS \citep{Grogin2011, Koekemoer2011} fields (AEGIS; \citealt{Davis2007}, COSMOS; \citealt{Scoville2007}, and GOODS-N; \citealt{Giavalisco2004}). Similarly, targets for the MOSDEF survey were selected from the multi-wavelength photometric and spectroscopic catalogues of the 3D-HST survey \citep{Brammer2012, Skelton2014}. 

As with the LEGA-C data, we classify galaxies in the MOSDEF survey as clumpy or nonclumpy using the HST photometric imaging products from the 3D-HST catalogue. In particular, we use the available HST imaging, spanning F475W to F160W, to construct rest-frame $U$-band brightness maps, following the procedures described in \S\ref{sec:ch3_method}. These images have an estimated AB magnitude depth of approximately 26-27. 
Finally, we note that the imaging products used for targets in the LEGA-C and MOSDEF surveys mentioned in this paper are solely for classifying galaxy morphology. We do not attempt to extract stellar properties of clumps, either due to these having limited spatial angular resolution (in the case of LEGA-C), or lack of sufficient filter coverage for SED fitting (in the case of MOSDEF). 

\subsection{Galaxy Sample}
We selected a sample of galaxies from the different surveys using several criteria:
\begin{enumerate}
    \item A $\mathrm{S/N}\geq3$ for $\oii$, $\oiiiA$, $\oiiiB$, and $\hb$ to ensure a robust estimate of the gas-phase metallicity for the LEGA-C sample. We additionally required $\mathrm{S/N}\geq3$ for $H\gamma$. $H\gamma$ was used to correct for dust based on the Balmer decrement when inferring gas-phase metallicities. 
    \item A $\mathrm{S/N}\geq3$ for $\ha$ and $\nii$ to ensure a robust estimate of the metallicity in both the MOSDEF and CANUCS samples. No dust correction was applied since $\ha$ and $\nii$ are close in wavelength.
    \item We excluded emission lines strongly affected by skylines in both the MOSDEF and CANUCS samples. For CANUCS, this was done by visual inspection, while for MOSDEF we used their skyline flag criterion, $\texttt{SLFLAG}>0.2$. 
    \item Finally, potential active galactic nuclei (AGNs) were excluded to avoid contamination of emission line ratios and derived metallicities.
    We excluded galaxies with \texttt{FLAG\_SPEC} \ensuremath{\geq 1} \citep{vandelWel2021} and galaxies flagged as mass-excitation AGNs using the cut from \cite{Juneau2011} for the LEGA-C galaxies. For MOSDEF and CANUCS, we excluded galaxies with $\log({\nii}/{\ha}) > -0.4$ \citep{Fernandes2010}.
\end{enumerate}

\begin{figure}[!t]
    \centering
    \includegraphics[width=\linewidth]{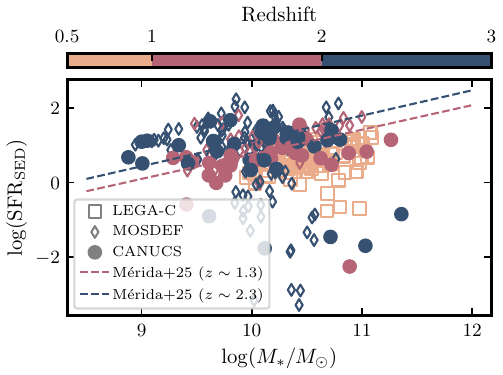}
    \caption{
    The mass and SFR distribution of galaxies from LEGA-C, MOSDEF and CANUCS, color coded by redshift. Both stellar mass and SFR are derived from \textsc{Dense Basis}. 
    The star-forming main sequences from \cite{Merida2025} are shown as the dashed lines. We do not show a star-forming main sequence for the LEGA-C sample as \cite{Merida2025} does not extend below $z=1$.
    }
    \label{fig:ch3_sample_selection}
\end{figure}

These criteria yielded a sample of robust emission lines for 94 LEGA-C, 125 MOSDEF, and {\nmosfiresamp} CANUCS galaxies. Figure \ref{fig:ch3_sample_selection} shows the distribution of galaxies based on their stellar masses and star formation rates. These were derived by fitting the SED using \textsc{Dense Basis} \citep{Iyer2017}. For the LEGA-C and MOSDEF galaxies, we also adopted \textsc{Dense Basis} to model the SED using the photometric catalogue from \cite{Muzzin2013} and \cite{Skelton2014} with the spectroscopic redshift.

\section{Methodology} \label{sec:ch3_method}

\subsection{Finding clumpy galaxies} \label{sec:find_clumps}

Our approach to determine if a galaxy is clumpy or nonclumpy is as follows. We began by spatially binning pixels together to achieve sufficient signal-to-noise (S/N) for SED fitting (see \S \ref{sec:ch3_binning}). We then fitted the SED of each bin to reconstruct the rest-frame $U$ (${\sim}3600$ \r{A}) surface brightness map (\S \ref{sec:ch3_Umap}). Galaxies were then classified as clumpy or nonclumpy based on these maps in \S \ref{sec:ch3_classification}. All classification and inferred properties were derived from PSF-matched images. Specifically, CANUCS data used PSF-matched JWST imaging, MOSDEF with PSF-matched imaging to HST WFC3 F160W, and LEGA-C with PSF-matched ground-based deconvolved image (see \citealt{Sok2022}). 

\subsubsection{Spatially binning pixels} \label{sec:ch3_binning}
Fitting the spectral energy distribution of galaxies in an unresolved manner has been shown to lead to overestimation of stellar mass (e.g., \citealt{Sorba2015, Sorba2018}). To mitigate this, we applied a spatially resolved SED fitting technique on PSF-matched imaging to obtain robust estimates of the physical properties of star-forming clumps and their host galaxies.
However, since individual pixels often have signal-to-noise ratios (S/N) that are unsuitable for SED fitting, we first spatially binned pixels together to improve the S/N while retaining some spatial information. This adaptive binning step was performed using the Voronoi tessellation technique (see \texttt{vorbin}, \citealt{Cappellari2003}).

We required each Voronoi bin to achieve a minimum S/N of 5. For each survey, we selected the image probing the reddest rest-frame wavelength to define the bins. This ended up being the deconvolved $K_s$ for LEGA-C, F160W for MOSDEF and F444W for CANUCS. The choice to use a redder rest-frame NIR filter was motivated by the fact that it closely traces the underlying stellar mass (e.g., \citealt{Wuyts2012, Sok2022}). However, this choice did not affect the identification of clumps, since we redistributed the inferred rest $U$ fluxes based on the observed rest UV images after fitting the binned SEDs (Section~\ref{sec:ch3_Umap}). This procedure preserved the UV morphology when mapping the fitted quantities back to the native pixel grid, under the assumption that all pixels within a given bin shared the same colours.

\subsubsection{Creating $U_\mathrm{rest}$ maps} \label{sec:ch3_Umap}
For each galaxy, we then calculated the fluxes in each Voronoi bin for each filter. 
We determined the flux in each bin by summing up its pixel values. The flux uncertainty was estimated by the quadrature sum of the statistical and systematic uncertainties, with the statistical uncertainty being the aperture flux uncertainty, while the systematic uncertainty accounted for potential biases during data reduction and calibration. In practice, the flux uncertainty for a given Voronoi bin was assumed to be the same as the flux uncertainty within a circular aperture encompassing the same area. The systematic uncertainty was considered to be a multiplicative factor of 5\% of the flux. This is similar to what is typically done in SED fitting (e.g., see \citealt{Abdurrouf2021}).

We used $\textsc{eazy-py}$ \citep{Brammer2008} to infer rest-frame $U$ luminosity ($U_\mathrm{rest}$) for each bin, adopting the recalibrated rest-frame $U$ filter from \cite{Apellaniz2006}. We set the redshift to each galaxy's spectroscopic redshift. As binning degrades angular resolution, particularly for coarser bins at the outskirts of galaxies where S/N is lowest, we recovered spatial resolution by ``dezonifying" each map \citep{Fernandes2013}. With this approach, the $U_\mathrm{rest, bin}$ for each pixel $i$ within a Voronoi bin were redistributed as, 
\begin{equation}
U^\prime_i = U_\mathrm{rest, bin} \frac{F_i}{F_\mathrm{bin}},
\end{equation}
\noindent where $F_i$ denotes the flux value at that pixel $i$, and $F_\mathrm{bin}$ is the total flux of the Voronoi bin. This preserved the total value for each bin while reintroducing spatial resolution. 

\subsubsection{Clumpy/nonclumpy classification} \label{sec:ch3_classification}

In this Section, we describe how galaxies were classified as clumpy and nonclumpy. The classification was based on the $U_\mathrm{rest}$ surface brightness map of the galaxy. To this end, we constructed a normalized light profile from the $U_\mathrm{rest}$ map using the prescription described in \cite{Wuyts2012}. We refer the reader to their paper and to \cite{Sok2022} for a detailed explanation of the methodology. Briefly, galaxies with no clumps (i.e., a smooth disk) have a smooth normalized $U_\mathrm{rest}$ light profile, decreasing in intensity with galactocentric radius. Since clumps are regions of enhanced star formation, clumps appear as a bump on an otherwise smooth normalized light profile in the $U_\mathrm{rest}$. 

\begin{figure*}
    \centering
    \includegraphics[width=\linewidth]{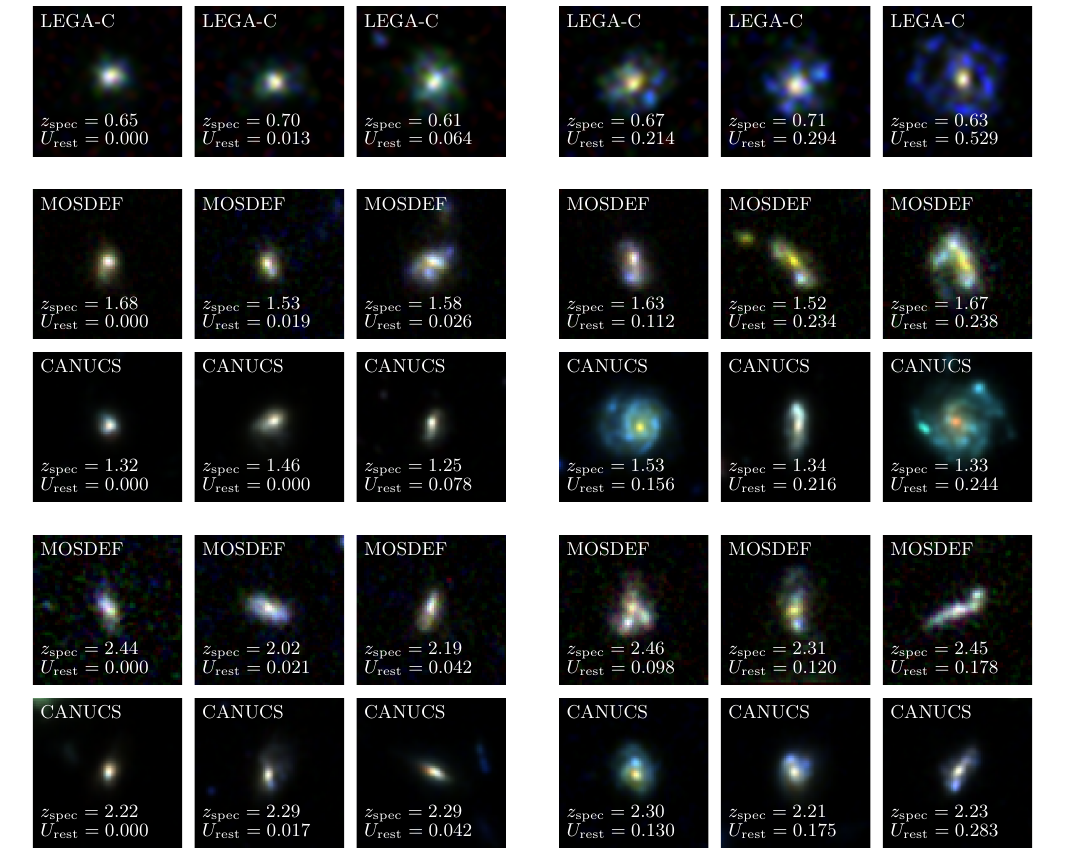}
    \caption{Examples of the classification of clumpy and nonclumpy galaxies. We show galaxies with increasing fractional $U_\mathrm{rest}$ luminosity from left to right, with the three left- and right-most galaxies being classified as nonclumpy and clumpy, respectively. Each row further separates galaxies by the different redshift intervals sampled by the different surveys. Each cutout is $3.6^{\prime\prime}\times3.6^{\prime\prime}$. }
    \label{fig:clumpy_examples}
\end{figure*}

The construction of the normalized light profile began with first defining elliptical parameters for each galaxy. The centre of mass of the galaxies was adopted from the stellar mass map, assuming it traces the gravitational potential. An elliptical aperture was then used to obtain the curve of growth centred on the centre of mass. A half-light radius $R_e$ was calculated as the effective radius at which half the light of the galaxy is contained within that radius. We then calculated the mean $U_\mathrm{rest}$ surface brightness within the half-light brightness as $\Sigma_e$. The galactocentric distance of each pixel within the $U_\mathrm{rest}$ brightness map was then normalized by the half-light radius, while the $U_\mathrm{rest}$ flux of each pixel was normalized by $\Sigma_e$. This renormalization mapped each pixel into a new parameter space, where \cite{Wuyts2012} defined a region within this parameter space to select clumpy pixels from the $U_\mathrm{rest}$ brightness map. The dividing line for the clumpy regime was determined by visually analyzing a number of normalized maps from different galaxies. In this paper, clumpy galaxies are defined as those galaxies with clumps contributing to at least $8\%$ of the total $U_\mathrm{rest}$ luminosity. This value is broadly consistent with other studies that identified clumps and clumpy galaxies (e.g., \citealt{Guo2015, Wuyts2012, Sok2025}). 
Examples of the classification of galaxies are shown in Figure \ref{fig:clumpy_examples}.

\subsection{Properties of Star-Forming Clumps}

For star-forming clumps in galaxies located in the CANUCS fields, we made use of the available deep medium- and broad-band JWST imaging to extract their physical properties, including stellar masses, star formation rates and ages. 
This section describes how individual clumps were identified within clumpy galaxies and how we determined their physical properties. We note that this sample is a subset of the clumpy galaxy sample presented in \cite{Sok2025b}. We therefore refer the reader to that paper for a more detailed description of the methodology; however, we briefly describe it here.

\subsubsection{Finding clumps in galaxies}
\begin{figure}
    \centering
    \includegraphics[width=\linewidth]{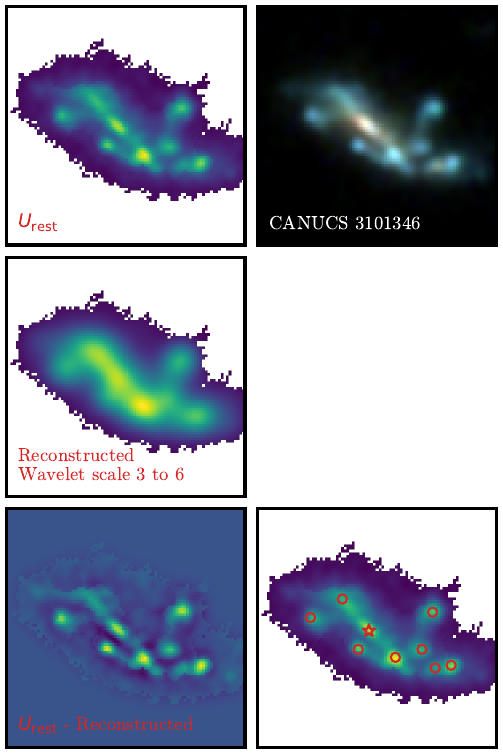}
    \caption{An example showing how clumps are identified. Clumps are detected based on the high-contrast image. This high-contrast image is obtained by subtracting the rest-frame bright map by its reconstructed smooth image (see \S \ref{sec:find_clumps}). The reconstructed image is shown in the middle left panel. The bottom left panel shows the high-contrast image, while the bottom right shows the detected clumps (circle) and bulge (star).}
    \label{fig:clump_find}
\end{figure}

Clumps were identified in a similar process as described in \cite{Kalita2025b}, where the $U_\mathrm{rest}$ image was passed through a high-pass filter to create a high-contrast image.
In practice, the low-pass image of the galaxy was determined by decomposing the image into six different wavelets using \textsc{scarlet} \citep{Melchior2018}. Each wavelet scale probed a different spatial scale. We obtained a smooth version of the image by setting the first and second wavelet scale as zero. Subtracting the image from its smoothed image resulted in an image containing only smaller-scale structures. Clump candidates were then identified using the peak-finding algorithm \texttt{peak\_local\_max} from the Python package \textsc{scikit-image} \citep{vanderWalt2014}.
We set a minimum $S/N > 3$ within an aperture of $0^{\prime\prime}.2$ to remove spurious detection of clumps. Figure \ref{fig:clump_find} illustrates this and highlights the detected clumps for a clumpy galaxy. 

\subsubsection{Inferring physical properties from SED}
Since star-forming clumps are generally superimposed within the disk of their host galaxies, background subtraction is typically performed to mitigate flux contamination from the disk. In previous studies, disk subtraction was commonly performed using aperture photometry, where the disk background flux is estimated from an annulus aperture (e.g., \citealt{Guo2018, Sattari2023}). We followed a similar methodology to subtract the disk. Specifically, we used an aperture size of $0^{\prime\prime}.2$ centred on the clumps, with an annuli aperture of $\sim0^{\prime\prime}.3-0^{\prime\prime}.35$.

The background-subtracted flux was then obtained for each filter, and the SED was fitted using \textsc{Dense Basis} \citep{Iyer2017, Iyer2019} to extract the physical properties of clumps. The star formation history (SFH) in \textsc{Dense Basis} is described by a several parameters, including $M_*$, SFR, ${t_\mathrm{X}}$). Here, we defined ${t_\mathrm{X}}$ as a set of three lookback times describing when the galaxy formed equally spaced quantiles of its total mass  (i.e., $t_{25}$, $t_{50}$, $t_{75}$). We adopted a Chabrier initial mass function \citep{Chabrier2003} and Calzetti attenuation law \citep{Calzetti2000} to model the stellar populations. The priors included flat priors for $\log(M_*)$ with range of (6,12),  $\log(\mathrm{sSFR})$ with range of (-14,-7),  $\log(\mathrm{Z/Z_\odot})$ with range of (-1.5,0.25), and $A_V$ with range of (0,4). 
\textsc{Dense Basis} performs SED fitting using an amortized brute-force Bayesian approach, in which an atlas of model SEDs is precomputed from random draws of parameters sampled from the prior distributions. 

We extracted the stellar masses, star formation rate (averaged over a lookback time of 100 Myr), and the age of the clumps. Here, the age of the clumps was determined as the lookback time at which $75\%$ of its total stellar mass has formed, based on the inferred star formation history. This definition of age is similar to the mass-weighted age, but is more sensitive to recent star formation episodes.

\section{Gas-Phase Metallicity} \label{sec:ch3_results}

This Section describes how we infer gas-phase metallicity from strong emission lines across different surveys. 
There exist several empirically/theoretically-calibrated metallicity diagnostics using strong emission line ratio (for a review see \citealt{Kewley2019}).
In this work, we utilized both the $R23$ and $N2$ indices to infer gas-phase metallicity. 

For the LEGA-C samples, metallicities were derived using the $R23$ strong line calibration from \cite{Kobulnicky2004}. The $R23$ index requires multiple strong emission lines to obtain the relevant flux ratios to infer gas-phase metallicities;

\begin{equation}
    R23 = \frac{\oii + \oiiiA + \oiiiB}{\hb},
\end{equation}
\begin{equation}
    O32 = \frac{\oiiiA + \oiiiB}{\oii}.
\end{equation}
The $R23$ index is sensitive to the ionization parameter, but this sensitivity can be corrected for using the $O32$ index. 
The typical intrinsic scatter for an individual metallicity measurement is ${\sim}0.1$ dex \citep{Kobulnicky2004}.

For the MOSDEF and CANUCS samples, we used the $N2$ index to infer metallicity.
The $N2$ index is commonly used in higher-redshift observation due to the close proximity of $\ha$ and $\nii$, and is defined by the line ratio, 
\begin{equation}
    N2 = \frac{\nii}{\ha}.
\end{equation} 
To this end, we followed the line calibration from \cite{Pettini2004}, which is given as, 
\begin{equation}
    12+\log(\mathrm{O/H}) = 8.9 + 0.57 \times N2.
\end{equation}
\noindent The intrinsic scatter of the relation is approximately $0.18$ dex.

We note that different line diagnostics provide different gas-phase metallicities. While there exist several ways to calibrate one line diagnostic to another \citep[e.g.,][]{Kewley2008, Teimoorinia2021}, we do not apply any such calibration in this paper, as the primary focus here is on the relative metallicity offset between clumpy and nonclumpy galaxies. 

\section{Do clumpy galaxies have lower metallicity compared to the MZR?} \label{sec:ch3_metallicity_analyses}


In previous works, we found that clumps have lower metallicity than their host disks, and that clumpy galaxies have lower metallicity than nonclumpy galaxies \citep[e.g.,][]{Sok2025, Estrada2025}. In particular, we showed that for a controlled sample of galaxies with similar masses, star formation rates and colours, clumpy galaxies are systematically more metal-diluted than their nonclumpy counterpart \citep{Sok2025}.
In this Section, we further investigate the metallicity offset between clumpy and nonclumpy galaxies across a broader sample.
We remind the reader that clumpy galaxies are defined as galaxies with star-forming clumps that contribute to at least 8\% of the total $U_\mathrm{rest}$ luminosity of the host galaxy. 

\begin{figure*}
    \centering
    \includegraphics[width=\linewidth]{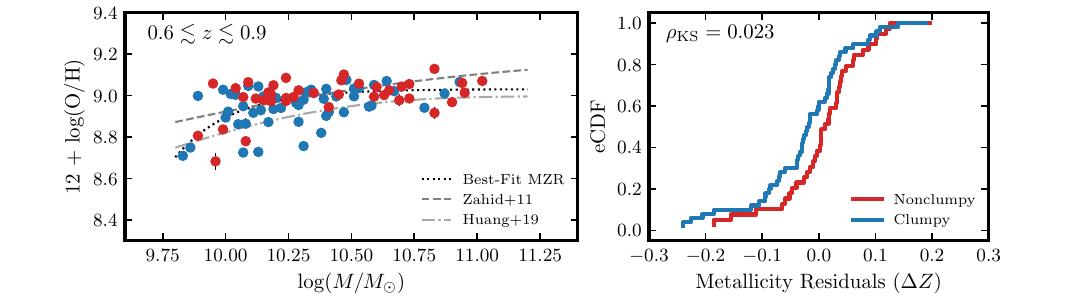}
    \caption{The mass-metallicity relation for the LEGA-C sample using the $R23$ index. The dotted line shows the best-fit mass-metallicity relation. The right panel shows the empirical cumulative distribution function (eCDF) of the metallicity residual, defined to be $\Delta Z = Z - \mathrm{MZR}$. We find that the eCDF of clumpy galaxies is generally shifted to lower $\Delta Z$ values compared to nonclumpy galaxies, indicating that clumpy galaxies have metallicity lower than their nonclumpy counterparts.}
    \label{fig:mzr_legac}
\end{figure*}

In these analyses, we did not apply any conversion factor between the $R23$ and $N2$ indices as we are mainly interested in the relative offset in gas-phase metallicity between clumpy and nonclumpy galaxies for each sample/redshift bin. We first examine the metallicity offset in the LEGA-C sample.
The left panel of Figure \ref{fig:mzr_legac} shows the gas-phase metallicity inferred from the $R23$ index. We also plot the mass-metallicity relation from \cite{Zahid2011} and \cite{Huang2019}, which inferred metallicity using the same line diagnostic for galaxies at a similar redshift. In general, we found that our metallicity range is consistent with previous measurements. 

Following our previous analyses in \cite{Sok2025}, we also show the metallicity offset between clumpy and nonclumpy galaxies as the distribution of $\Delta Z$, defined to be the offset between individual metallicity values to the mass-metallicity relation. To this end, we fitted the mass-metallicity relation using the parametrization given in \cite{Curti2020},
\begin{equation}
    12 + \log(\mathrm{O/H}) = Z_0 - \frac{\gamma}{\beta}\log \Big[ 1+ \Big(\frac{M}{M_0}\Big)^{-\beta} \Big] , 
\end{equation}
\noindent where $Z_0$ represents the asymptotic metallicity at the high-mass end, and $M_0$ is the characteristic turnover mass above which the relation begins to flatten. At masses below $M_0$, the MZR follows a power-law behaviour with slope $\gamma$, while $\beta$ controls the sharpness of the transition between these regimes.
The right panels of Figure \ref{fig:mzr_legac} therefore show the empirical cumulative distribution function (eCDF) constructed from $\Delta Z$ for both clumpy and nonclumpy galaxies. We found that the eCDF of $\Delta Z$ for nonclumpy galaxies is generally shifted to the right of clumpy galaxies, suggesting that the typical values of $\Delta Z$ for nonclumpy galaxies are typically higher compared to clumpy galaxies. 

\begin{figure*}[!t]
    \centering
    \includegraphics[width=\linewidth]{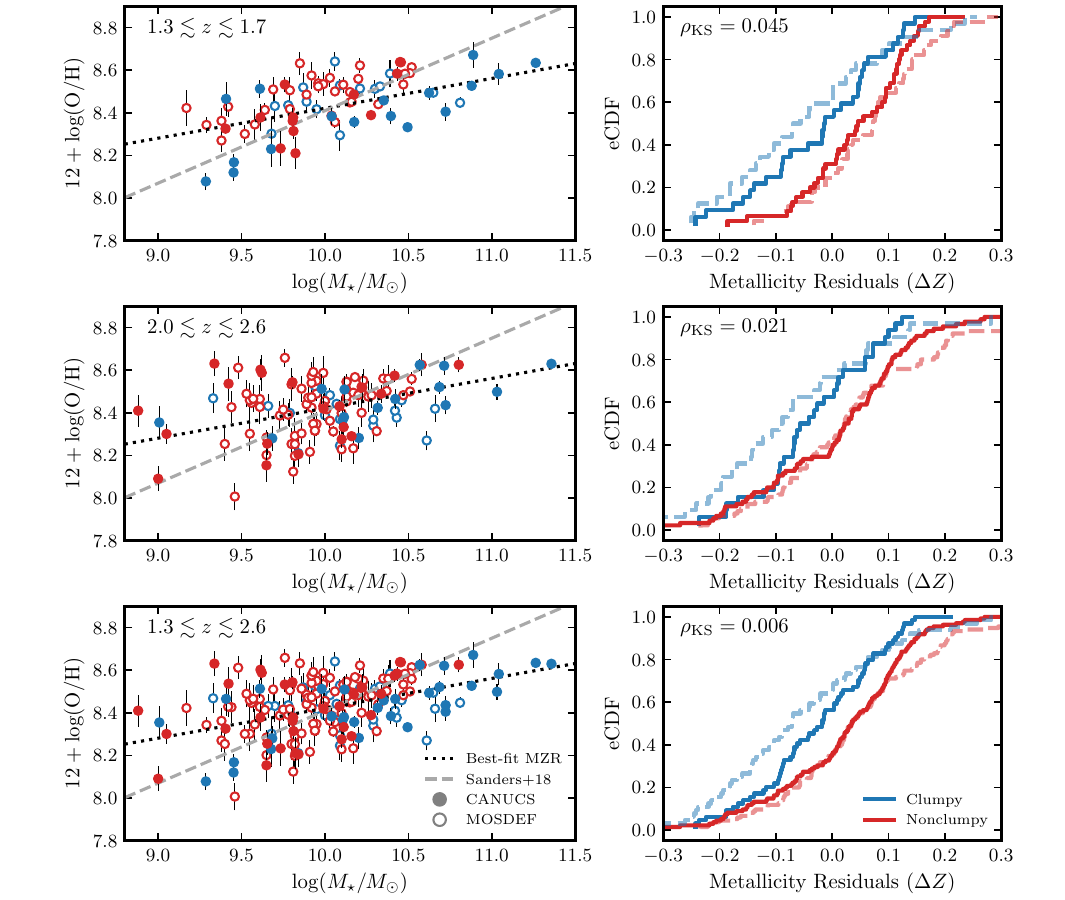}
    \caption{Similar to Figure \ref{fig:mzr_legac}, but for the MOSDEF and CANUCS samples. We show the observed mass-metallicity relation at different redshift bins for each row in the left panels, while the right panels show the eCDF. The best-fit mass-metallicity relation is determined by combining all galaxies between $1.3<z<2.6$ as shown in the bottom left panel. The dashed eCDFs in the right panels are computed using the mass-metallicity relation from \cite{Sanders2018}.}
    \label{fig:mzr}
\end{figure*}

Similarly, the left panels of Figure \ref{fig:mzr} show the distribution of the gas-phase metallicity inferred from the $N2$ index. We plotted the best-fitting MZR, obtained by combining the sample of galaxies with $1.3<z<2.6$. We also plotted the MZR from \cite{Sanders2018} to compare with our derived gas-phase metallicity. Their MZR was obtained from a sample of 260 SFGs in the MOSDEF sample at $z\sim2.3$. We note that the slope of our MZR is shallower compared to the MOSDEF survey. However, using their best-fit MZR does not affect the main result of this study (e.g., that is, clumpy galaxies have lower $\Delta Z$ as compared to nonclumpy galaxies, as shown by the dashed eCDFs in the right panel of Figure \ref{fig:mzr}). In addition, when calculating the $\Delta Z$ to our best-fit MZR, we found that the weighted mean difference between nonclumpy and clumpy galaxies is $0.05 \pm 0.03$ dex, where the uncertainty reflects both measurement errors and the intrinsic scatter of 0.18 dex from the N2 metallicity calibration. This weighted mean difference is $0.10 \pm 0.03$ dex when we use the MZR from \cite{Sanders2018}.

Finally, we performed a two-sample Kolmogorov-Smirnov (KS) test to determine whether the eCDF of $\Delta Z$ for nonclumpy galaxies is systematically lower than that of clumpy galaxies. The resulting p-values $\rho_{KS}$ are reported in the right panels of both Figure \ref{fig:mzr_legac} and \ref{fig:mzr}, where a value of $\rho_{\mathrm{KS}} < 0.05$ indicates less than a five percent probability that the observed difference in eCDFs arises by chance. In general, we found that $\rho_{\mathrm{KS}} < 0.05$, indicating that the two eCDFs are statistically different, and that clumpy galaxies intrinsically have lower metallicities than nonclumpy galaxies. These results are consistent with what we reported in \cite{Sok2025}.

We note here that the finite slit width of $0^{\prime\prime}.7$ introduces some complexity in interpreting the inferred metallicities. In particular, the MOSFIRE slit position angles were random in order to maximize object priorities rather than be aligned along the major axis as in \citet{Sok2025}. This means that the slit may not fully cover all clumps in some clumpy galaxies (for examples, see Figure \ref{fig:spectra_example}), introducing scatter to their inferred metallicities, while inferred metallicities for nonclumpy galaxies are more representative as the slit generally spatially covers them. Regardless of slit placement and width, we find that clumpy galaxies have systematically lower metallicities than nonclumpy galaxies even in the presence of this scatter, suggesting the offset is robust to these observational limitations. 

It is also important to consider the underlying galaxy sample for both clumpy and nonclumpy populations, and how it may influence the result. The fraction of clumpy galaxies depends on both stellar mass and SFR. Broadly speaking, low-mass systems are predominantly nonclumpy, while higher-mass systems are more likely to be clumpy (e.g., \citealt{Sok2025b}). The abundance of massive galaxies (above $10^{10.5}~M_\odot$) in the CANUCS data likely reflects selection biases, with massive clumpy galaxies prioritized in our observations. While these galaxies have lower metallicity as compared to the mass-metallicity relation, it is possible that nonclumpy galaxies at similar high masses also have low metallicity, but are underrepresented in our sample.
To assess whether these effects influence our results, we repeated the analysis within a restricted mass range of $9.4 < \log(M_*/M_\odot) < 10.6$, and randomly drew from the nonclumpy sample to match the mass distribution of the clumpy sample.  This gave a $\rho_{KS}=0.04$ for the combined CANUCS and MOSDEF sample between $1.3<z<2.6$.

\section{Stellar Population of Clumps} \label{sec:ch3_clump_properties}

As shown in the previous section, clumpy galaxies are generally metal-poor compared to nonclumpy galaxies. The recent work of \cite{Estrada2025} further showed that clump populations consist of those with metallicities similar to the host disk and those that are lower. The galaxies that host lower metallicity clumps are generally found to have lower metallicities compared to those galaxies that host clumps with no metallicity offsets. These observations suggest that clump formation, at least those with lower gas-phase metallicity, is the result of metal-poor gas accretion events.

An interesting question is how the properties of star-forming clumps differ between those above and below the mass-metallicity relation. The stellar masses and ages of clumps have been shown to provide insights into their origins (e.g., \citealt{Ribeiro2017, Zanella2019, Sok2025b}). In this section, we therefore examined clump properties as a function of their host galaxy’s offset from the MZR. Here, we omitted the LEGA-C samples because the deconvolved ground-based images have lower resolution. The following analyses therefore only consist of derived properties for the CANUCS samples, using both HST and JWST imaging. Similar to \cite{Sok2025b}, we only computed the stellar properties of clumps found in galaxies that are not highly magnified ($\mu < 2$) and that are not edge on (axial ratio of $q>0.2$). This resulted in a sample of 30 clumpy galaxies, with ${\sim}130$ clumps.

\begin{figure}[!t]
    \centering
    \includegraphics[width=\linewidth]{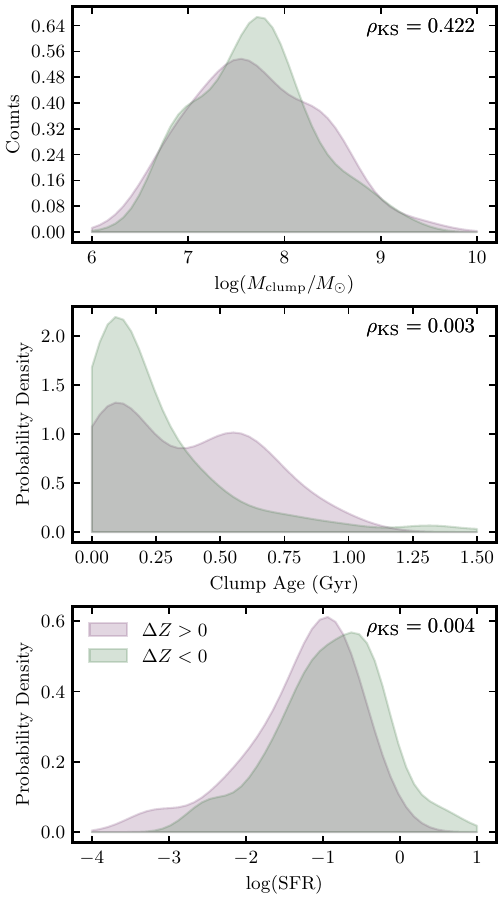}
    \caption{The kernel density estimate of clump masses, ages, and SED-derived SFR for clumps in galaxies with $\Delta Z > 0$ (purple) and $\Delta Z < 0$ (green).
    In general, we find that clumps in galaxies that lie below the mass-metallicity relation are younger and highly star-forming compared to clumps in galaxies that lie above the mass-metallicity relation.}
    \label{fig:clump_populations}
\end{figure}

We divided the sample into clumps in galaxies above the MZR ($\Delta Z > 0$) and those below it ($\Delta Z < 0$). In Figure \ref{fig:clump_populations}, we present the distributions of clump stellar masses, ages and SED-derived SFR for these two samples. To improve statistics, we combined the two redshift bins, resulting in a sample of approximately $120$ clumps between $1.3<z<2.6$. Clumps in host galaxies that lie below the MZR tend to be younger than those in hosts above the MZR. The age distribution of clumps in hosts above the MZR also appears to be bimodal, possibly suggesting episodic gas accretion. Additionally, clumps in galaxies below the MZR have higher SFRs compared to those above the MZR.

To quantify these differences, we performed the KS test to determine whether the mass and age distributions of clumps in galaxies below the MZR are smaller and younger than those above the MZR. 
For clump masses, the KS test returned a p-value greater than 0.05, suggesting that the mass distributions are not statistically different. In contrast, both the age and SFR distributions yielded p-values less than 0.05, implying that clumps below the MZR are statistically younger and exhibit more active star formation than those above the relation. While these trends are consistent with a picture in which metal-poor gas accretion drives the formation of younger, more actively star-forming clumps, larger samples will be needed to place more robust constraints on the physical origin of clump populations.


\section{Discussion} \label{sec:ch3_discussion}


The mass-metallicity relation (and similarly, the fundamental metallicity relation) provides a framework for connecting galaxy evolution to physical processes such as star formation, the infall of metal-poor gas, and the outflow of enriched material. 
In particular, scatter in the gas-phase metallicity around the relation is often interpreted as being driven by the accretion of metal-poor gas and/or mergers. 
In this Section, we discuss whether mergers or accretion from cosmological streams drive the resulting metallicity offset and clump formation.

Generally, metal-poor gas accretion events lead to the dilution of the gas-phase metallicity \citep[e.g.,][]{Sanchez2014, Ceverino2016, Sok2025, Estrada2025}. 
Similarly, mergers can also affect the scatter of the mass-metallicity relation. In interacting galaxies, the metallicity dilution and enrichment can be driven by the inflow of metal-poor or metal-rich gas already present within the merging galaxies. In the former case, metallicity dilution of around 0.1 dex is often attributed to major mergers (with a mass-ratio larger than $\mu > 1/3$), while weaker amplitudes in metallicity offsets are expected for minor mergers \citep{Bustamante2018, Bustamante2020}. This interpretation is often inferred from observing an increase in metal dilution with respect to decreasing projected separation between galaxy pairs \citep[e.g.,][]{Omori2022}. However, the outcome of the enrichment process likely depends on the initial metallicities of the interacting galaxies. For example, \cite{Michel2008} found that the secondary member of minor mergers exhibits metallicities $\sim$0.2 dex higher compared to the mass–metallicity relation, possibly due to accretion of enriched gas from the more massive companion. 


To determine whether clumps in our sample are formed by recent mergers, we identified potential recent mergers using a statistical morphological classification based on the Gini ($G$) versus $M_{20}$ (e.g., \citealt{Lotz2004}). We computed these nonparametric morphological quantities based on the stellar mass maps of galaxies using \textsc{statsmorph} \citep{Rodriguez2019}, where mergers have a merger statistic that satisfies $S_\mathrm{merger}(G, M_{20}) > 0$. To this end, we adopted the updated merger statistics equation from \citet{Tan2024}, which were calibrated against visual classifications of galaxies in the CANUCS fields to identify mergers between $0.3<z<5$. Out of the {\nmosfiresamp} galaxies in CANUCS with robust metallicities, only 30 of them were classified as clumpy. However, we find that only one of them is classified as a merger.


We emphasize that the merger classification based solely on $S_\mathrm{merger}(G, M_{20})$ is generally not complete (e.g., \citealt{Faisst2025}). The observability timescales for mergers is fairly short, with the merger observability timescale based on G–M20 morphology estimated to be $0.2$–$0.5$ Gyr \citep{Lotz2008, Lotz2010}. In contrast, the metallicity dilution phase can persist for ${\sim}0.5-1.5$ Gyr for a 1:1 merger scenario \citep{Montuori2010, Gronnow2015, Bustamante2018}. However, since the ages of our clumps in metal-diluted galaxies are generally less than $500$ Myr (e.g., Figure \ref{fig:clump_populations}) and this is comparable to the merger observability timescale, it suggests that merger signatures should be detectable for these galaxies if mergers were the cause for the observed metallicity offsets. This is not the case in our sample.

Nevertheless, resolved kinematic observations are needed to identify merger systems. While such information is not available for the current sample, we showed in our previous study of ${<}20$ star-forming galaxies at $z\sim0.8$ that the kinematics of both clumpy and nonclumpy galaxies are broadly similar and consistent with rotation-dominated systems \citep{Sok2025}. Mergers-driven clumps may be more important at $z>6$ (e.g., \citealt{Vega2025}), or in an overdense environment \citep{Sattari2023}. Figure \ref{fig:clump_populations} is therefore consistent with a scenario in which the accretion of metal-poor gas drives metallicity dilution. Star formation then occurs within the accreted gas, as indicated by the presence of younger, highly star-forming clumps in galaxies that lie below the mass–metallicity relation. 



\section{Conclusion} \label{sec:ch3_conclusion}

The origin of star-forming clumps is often linked to violent disk instabilities, where global instabilities in disks of high-$z$ galaxies arise from the continuous accretion of metal-poor gas. Until recently, direct observational constraints on the link between clump formation and physical processes, such as gas inflow, have been limited. Recently, studies have reported evidence for gas inflow and metallicity dilution in clumpy galaxies, albeit for a small sample of ${<}30$ galaxies \citep{Sok2025, Estrada2025}. 

In this study, we compile spectroscopic and photometric data from the LEGA-C and MOSDEF surveys, combined with new MOSFIRE observations of galaxies in the CANUCS fields, to investigate the relationship between gas-phase metallicity and galaxy morphology. We infer metallicities using the $R23$ and $N2$ indices in a sample of $\sim300$ galaxies between $0.6<z<2.6$. The main findings are summarized as follows: 
\begin{itemize}
    \item Across all redshift bins, clumpy galaxies exhibit systematically lower metallicities compared to nonclumpy galaxies, relative to the mass-metallicity relation (MZR)
    \item For clumps in the CANUCS field, we performed SED fitting to extract their mass, SFR, and age. We found no significant mass differences between clumps in galaxies lying above or below the mass–metallicity relation (MZR).
    \item However, galaxies with metallicity below the MZR generally host younger and higher star-forming clumps, compared to galaxies with clumps that lie above the MZR.
    \item Based on the $S_\mathrm{merger}(G, M_{20})$ classification of the stellar mass maps, the majority of galaxies are inconsistent with being classified as a merger. This indicates that metallicity offsets in clumpy galaxies are more likely driven by metal-poor gas accretion events than by mergers.
\end{itemize}
Together, the results of this study support a scenario in which clump formation occurs due to gas accretion events, as reflected by the dilution of metals in clumpy galaxies. Furthermore, clump populations in galaxies with metallicity above the MZR are generally older and show less star formation than those below the MZR, suggesting possibly episodic accretion events. Future work incorporating resolved kinematic studies will be essential to disentangle the relative roles of gas inflow and mergers in driving metallicity dilution and clump formation.


\begin{acknowledgments}
Some of the data presented herein were obtained at Keck Observatory, which is a private 501(c)3 non-profit organization operated as a scientific partnership among the California Institute of Technology, the University of California, and the National Aeronautics and Space Administration. The Observatory was made possible by the generous financial support of the W. M. Keck Foundation.

The authors wish to recognize and acknowledge the very significant cultural role and reverence that the summit of Maunakea has always had within the Native Hawaiian community. We are most fortunate to have the opportunity to conduct observations from this mountain.

This research was supported by grants 23JWGO2A13, 23JWGO2B15 and  24JWGO2A04 from the Canadian Space Agency (CSA) and funding from the Natural Sciences and Engineering Research Council of Canada (NSERC). AM acknowledges support from the Yavin Family Fund.

GW gratefully acknowledges support from the National Science Foundation through grant AST-2347348. 

MB and NM acknowledge support from the ERC Grant FIRSTLIGHT and the Slovenian
National Research Agency ARRS through grants N1-0238, P1-0188. MB acknowledge support from the European Space Agency through Prodex Experiment Arrangement No. 4000146646.

GN acknowledges support by the Canadian Space Agency under a contract with NRC Herzberg Astronomy and Astrophysics.

DM acknowledges generous support from the Leonard and Jane Holmes Bernstein Professorship in Evolutionary Science. Support for programs JWST-GO-03362 and JWST-GO-05890, provided through a grant from the STScI under NASA contract NAS5-03127, is acknowledged.
\end{acknowledgments}

\bibliography{article}{}
\bibliographystyle{aasjournalv7}



\end{document}